# Assessment of the GCT prototype's optical system implementation and other key performances for the Cherenkov Telescope Array

**A. Dmytriiev[1,a] , L. Dangeon[b], G. Fasola[b], H. Sol[a], A. Zech[a], J. Gironnet[c], O. Le Blanc[b], J.P. Amans[b], G. Buchholtz[c], J.L. Dournaux[b], F. de Frondat[b], D. Horville[b], J.M. Huet[b], I. Jégouzo[b], P. Laporte[b] for the GCT collaboration and the CTA consortium[d]**

[a] *Observatoire de Paris, CNRS, Université Paris Diderot, PSL Research University, LUTH,
5 Place J. Janssen, Meudon, France*
[b] *Observatoire de Paris, CNRS, PSL Research University, GEPI, 5 Place J. Janssen, Meudon, France*
[c] *CNRS, DT-INSU, 1 Place Aristide Briand, 92190 Meudon, France*
[d] *http://www.cta-observatory.org , for consortium list see PoS(ICRC2019)1177*

The Cherenkov Telescope Array (CTA) project, led by an international collaboration of institutes, aims to create the world's largest next generation observatory for Very High Energy (VHE) gamma-ray astronomy. It will be devoted to observations in a wide band of energy, from a few tens of GeV to a few hundreds of TeV with Large, Medium and Small-sized telescopes.

The Small-Size Telescopes (SSTs) are dedicated to the highest energy range above a few TeV and up to 300 TeV. GCT is an imaging atmospheric Cherenkov telescope (IACT) proposed for the subarray of about 70 SSTs to be installed on the Southern site of CTA in Chile. The Observatory of Paris and the National Institute for Earth Sciences and Astronomy (INSU/CNRS) have developed the mechanical structure, mirrors (aspherical lightweight aluminium segments) and control system of the GCT. The GCT is based on a Schwarzschild-Couder (S-C) dual-mirror optical design which has the advantages, compared to the current IACTs, to offer a wide field of view (~ 9°) while decreasing the cost and volume (~ 9 m x 4 m x 6 m for ~ 11 tons) of the telescope structure, as well as the camera. The prototype (pGCT) has been installed at the Meudon's site of the Observatory of Paris and was the first S-C telescope and the first CTA prototype to record VHE events on-sky in November 2015.

After three years of intensive testing, pGCT has now been commissioned. This paper is a status report on the complete GCT telescope optical system and the performance it can provide for CTA.

---

[1]Speaker
E-mail: anton.dmytriiev@obspm.fr

**1. Introduction**

The GCT (Gamma-ray Cherenkov Telescope) is one of three designs proposed for the SST section of the CTA South array that will be installed in Chile. A prototype has been manufactured in 2014, installed and inaugurated on the Meudon site of the Paris Observatory in 2015. Contrarily to the current single-dish generation of IACT telescopes, the S-C design uses two mirrors to collect the Cherenkov light from high-energy air showers and project it onto the camera. The relevance of dual-mirror telescopes has been well established over centuries in astronomy. Such telescopes have rather compact mechanical structures and for a proper choice of mirror surfaces, aberrations of the primary can be canceled by those of the secondary, providing a better image quality over a larger field of view. The capabilities of dual-mirror systems and especially of the S-C design for ground-based gamma-ray astronomy have been recognized more recently [1]. Configured to correct spherical and coma aberrations and minimize astigmatism and vignetting, this type of design is isochronous and provides wide fields of view together with reduced plate scales. This design is thus directly compatible with small-size SiPM pixel cameras that do not require Winston cones, and can meet all needs of Cherenkov astronomy. The S-C design had never been built in astronomy before the advent of CTA, mainly because of the difficulty to manufacture aspherical non-conic mirrors, which is now overcome with present technology. Dual-mirror designs proposed to CTA are therefore the result of recent R&D and offer to the scientific communities a new type of astronomical telescopes for SSTs, which is also being implemented by the SCT team on a larger size-scale [2]. Dual-mirror telescopes will be particularly suitable for future upgrades of SiPM cameras with smaller pixel sizes. In the following sections, we provide first an overview of this novel telescope design, followed by a description of the prototyping and the current and expected performances.

**2. General design of the telescope structure and mirrors**

The mechanical structure has been designed with the aim to reduce weight and costs, and to ease shipping, assembly, operations and maintenance. The optical system has been chosen to optimize the point spread function (PSF) and the plate scale while keeping a maximum incident angle of 60° on the detector surface. The telescope can be equipped with both CHEC-like [5] and ASTRI-like [12] cameras. Optimisation across a range of field angles resulted in a GCT design with a 2 m diameter secondary mirror (M2), placed 3.56 m in front of a 4 m diameter primary mirror (M1). This creates a compact layout with a detector surface of radius of curvature 1.0 m, placed 0.51 m in front of M2. The focal length of the system is 2.283 m. Fig. 1 illustrates the final design, whilst the primary characteristics of the GCT design are listed in Table 1. The telescope structure itself is an altitude-azimuth design with a range in azimuth from − 90° to 450° and of up to 90° in elevation. The GCT Mechanical Assembly is mainly made of standard steel (E355) and contains three subsystems: (i) the Telescope Base which consists of a tower and its fixation to the foundation; (ii) the Altitude-Azimuth (alt-az) System (AAS) supported by the tower; (iii) the Optical Support Structure (OSS), which is attached to the AAS and holds the Optical Assembly. The M1 consists of six hexagonal lightweight aluminium mirror segments (alleviated rear structure). The M2 consists of a highly curved semi-monolithic surface. The Telescope Control System (TCS) hardware core is an industrial PC (*Beckhoff*); it hosts the main Programmable Logic Controller and the on-board control software. The link for control and command to and from the distant devices is limited for the structure to a single OPC UA server. A single field bus (*EtherCAT*) is used for all distributed inputs/outputs. A dedicated hardware module (*Beckhoff*) hosts the safety PLC connected to all related sensors and actuators via the same field bus. A proprietary network (*TransnET*) is required to manage the *ETEL* drives and torque motors in real-time via a PCI card (*UltimET*),

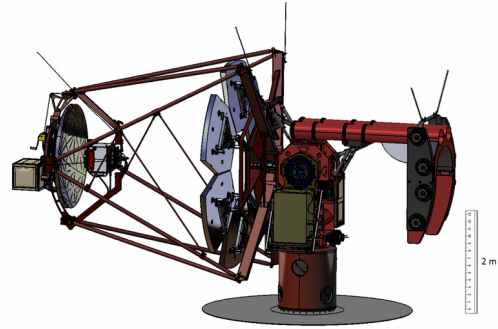

*Figure 1: CAD model of the final GCT design, with the segmented primary mirror to the right, the secondary mirror to the left and the camera in-between. A stick with length of 2 meters is shown for scale.*

set up and triggered in the Drive software and using the library provided by *ETEL*.

| Optical Parameters | | Mechanical Parameters | |
|---|---|---|---|
| Field of View *(depends on photosensor )* | 8.2° - 9.2° | M1 diameter | 4 m |
| Focal length | 2283 mm | M2 diameter | 2 m |
| focal length / diameter | 0.57 | Telescope size *(parking position)* | 4.1 m x 5.7 x 8.5 m |
| Plate scale | 39.6 mm/° | Telescope mass *(with CHEC camera)* | 10.8 tons |
| Throughput | > 60% | Distance M1 to M2 | 3.56 m |
| Focal plan radius | 1 m | Distance M2 to camera | 0.51 m |
| Effective mirror area *(corrected for shadowing)* | 7 m² (on-axis) 6.25 m² (4° off-axis) | | |
| PSF D80 on-axis | Aim: 3 mm (0.076°) | | |

*Table 1: Main parameters of the GCT structure. We characterise the point spread function (PSF) by the diameter of a circle containing 80% of energy (D80).*

The compactness of SST dual-mirror designs and especially the small dimensions of the GCT in "Parked" position provides direct access from the ground to all telescope sub-systems. This avoids work at height and is a great advantage in terms of safety but also time saving during assembly and operation. Design simplicity and maturity have been demonstrated during the prototyping phase described in the next section.

### 3. Prototyping

The GCT prototype (Fig. 2) recorded the first Cherenkov light on sky in November 2015 at the Paris Observatory in Meudon, just before its inauguration [3]. During a short campaign, the telescope was equipped with the CHEC-M camera, fully installed on the telescope structure in only two days and for the first time. Cosmic-ray showers were recorded despite a partially cloudy sky and high night sky background (NSB) due to the city lights from Paris. A second observation campaign with CHEC-M in March and in April 2017 in Meudon registered thousands of cosmic-ray showers under various conditions in agreement with the expected instrument performance [4]. These two campaigns provided a test bed to assess operational and maintenance procedures both for the telescope and the camera and proved the functionality of the S-C optics to record cosmic showers and of the CHEC camera for the use as an IACT camera in general. They allowed the involved teams to verify interfaces, to improve procedures, and to better understand the system variability and reliability. Data gathered during these campaigns have been proven useful in the development of the data-analysis chain and in understanding the levels of calibration that will be required for CTA [5].

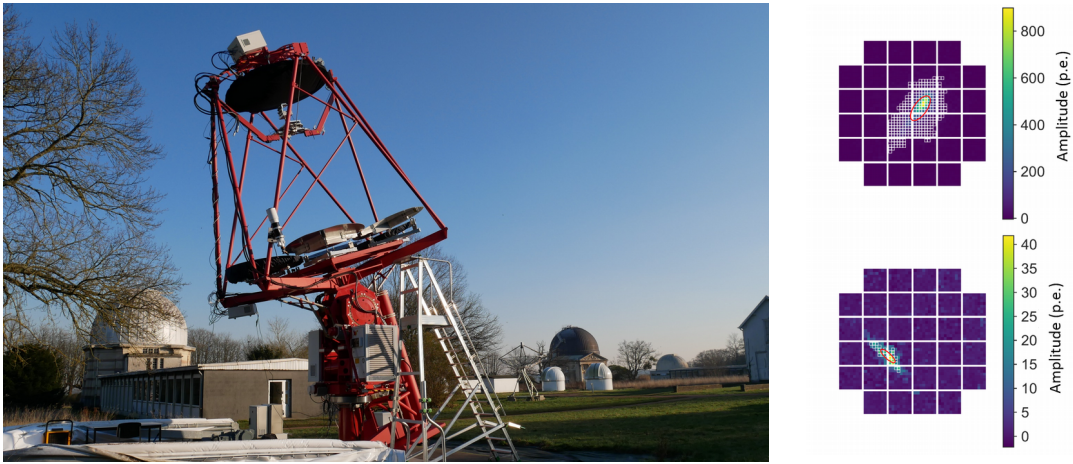

*Figure 2: Left: The GCT prototype and its open shelter at the Meudon site of the Paris Observatory during Cherenkov campaigns. The prototype was equipped with two M1 panels out of 6 and the MAPM-based camera of the CHEC team. Right: Examples of Cherenkov events detected by CHEC-M in Meudon. Credit: Observatoire de Paris*

Based on the knowledge acquired during the prototyping phase, the consolidation of the design for future telescopes has been carried out in order to further improve the performance, ease the manufacturing process or the maintenance and increase commonalities with other telescopes of CTA [6]. In parallel, tests were carried out in 2018 to enhance the aluminium mirrors surface quality by optimizing the manufacturing process, probing several options on flat 15 cm witness-samples, as well as equipping the GCT with two new M1 panels. Indeed, the choice of metallic technology for mirrors was driven by its advantages in terms of geometrical accuracy of the surface shape, easy maintenance, and lifetime superior to 30 years. But to benefit most from the high strength-to-weight ratio and other advantages, the mirrors are manufactured directly and continuously from an aluminium blank and their dimension, particularly of the M1 panels, represents the main challenge for their manufacturing (maximal dimension is their width of 1380 mm). The necessary trade-off between cost and quality made in 2014 for the secondary mirror and the two first primary segments of the prototype favoured the quality of machining and global mirror shape at the expense of the quality of the polishing and micro-roughness. That choice is reflected by the initial optical performance of the prototype which proves that the shape of the mirrors agrees with expectations, while the PSF is widened by their micro-roughness (see section 4). The witness-samples (cf. Fig. 3) from recent tests show a significant improvement of the surface quality and enabled to determine the best production steps: machining, lapping, Ni plating, polishing and optical coating. Die casting was also considered but proved not efficient to reach the desired surface quality (RMS micro-roughness Rq $\leq$ 7 nm) due to the structure of the casted aluminium. The process will be qualified upon delivery of the M1 panels by validating on-sky the PSF improvement before the end of the year.

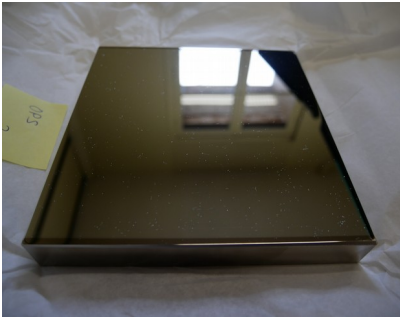

*Figure 3: Picture of one Al witness-sample of the recent tests.*

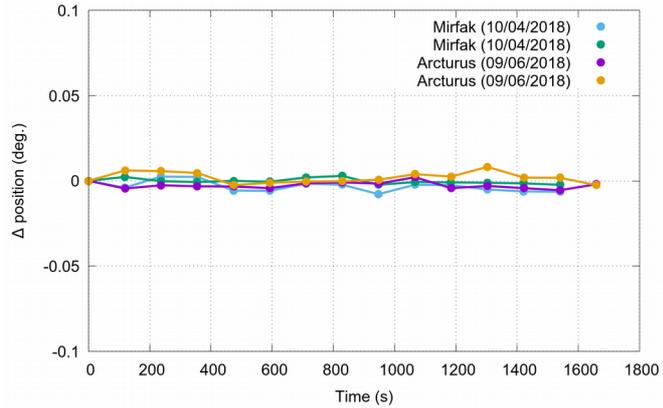

*Figure 4: Tracking accuracy of the GCT prototype during 30 min runs on both axes on two different sky positions and nights.*

## 4. Performance

### 4.1 Mechanical and control-command performance

The global geometry of the prototype mechanical structure equipped with metallic mirrors is found to be entirely compliant to the theoretical design, as shown by the experimental focal plan position, which is found within 1 mm of its theoretical value, and the on-axis plate scale of 39.57 mm/°, similar to the theoretical value of 39.6 mm/°. The stability of the structure and its experimental modal analysis (measured eigenmodes on the prototype) satisfies the normal modes specification and the optical displacement requirements [6, 7, 11]. The security systems and more especially the emergency stops have been tested several times during the different campaigns and the compliance tests, proving that the structure (and its pointing performance) suffer no damage. The tracking performance of pGCT is well below the required limit of 0.1 degrees on both axes. Examples of the GCT tracking accuracy are given in Fig. 4 for different weather conditions, while tracking a star with an ATIK CCD camera installed at the focal plane.

### 4.2 Optical performance

#### Prototype mirrors from 2014

As shown in Fig. 5, the PSF initially observed with the GCT prototype in Meudon appears degraded by scattering due to the high micro-roughness of the first mirrors installed in 2015 ($\sim$ 55 nm rms). Star image 1D and 2D profiles, obtained with the full M2 and one M1 segment, are well reproduced by double Gaussian fits, with a narrow and wide component. The PSF, which can be decomposed into two Gaussian functions, can be reproduced by light diffusion on both the M1 and M2 mirrors in a Gaussian distribution with standard deviation σ of 4.3 arc minutes. As proved with ROBAST simulations [9], and in agreement with theoretical arguments, the narrow component is due to the specular spot and to rays of light diffused mainly by the M2, while the large component is due to rays of light diffused by the M1 only, and to rays diffused twice, by the M1 and by the M2. In our starlight measurements, the narrow component has a D80 of 7.8 mm +/- 0.5 mm, which corresponds to 0.197° on sky. Observations of a sample of bright stars with an ATIK CCD camera at different elevation, on-axis, and off-axis at incident angle of 5° showed no significant variation of the D80 value of the narrow component, while the wide component was more variable. Those results are in agreement with ROBAST simulations (Fig. 6). The PSF being here dominated by the scattered light from both mirrors rather than the specular light flux, is not sensitive to the field angle.

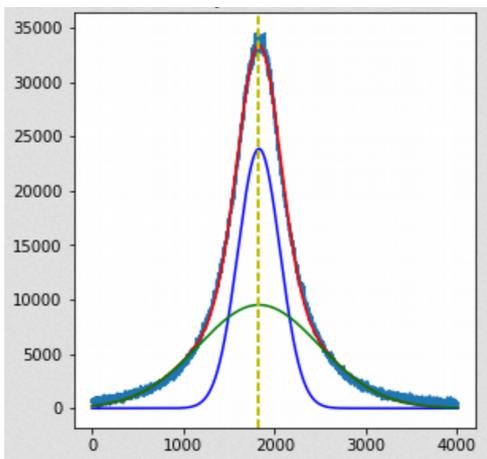

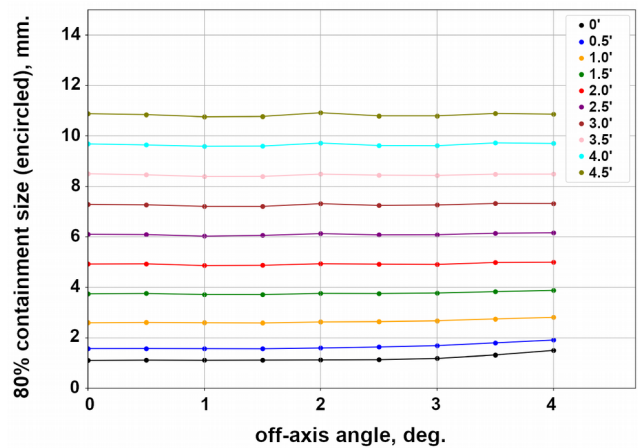

*Figure 5: Profile of the PSF from a bright star observed on-axis: single row crossing the image barycentre in blue, fitted with a sum of two Gaussian functions, shown in red. The individual Gaussians are shown with green and bright blue lines. (Units are ADC counts vs. pixel counts.)*

*Figure 6: ROBAST simulations of the impact of micro-roughness on the PSF over the required FoV (80% containment radius, reminder: plate scale is 39.6 mm/°).*

Images in optical broad-band filters were also obtained to test the behaviour of the PSF as a function of wavelength. The D80 values of both the narrow and wide components are found smaller with red than with blue filters, with a significant difference of about 20% . This provides an additional confirmation of the degradation of the PSF by scattering, which affects shorter wavelengths more strongly.

### Perspectives with the optimized GCT mirrors

While scattering effects add a wide diffuse component to the PSF, and even broadens the narrow component, modelling with ZEMAX OpticStudio © [10] of the performance using the large-scale surface characterization of the first 2014 mirrors of the prototype leads to estimates of about 2.2 mm for the specular D80 PSF on-axis (see Fig. 8). This confirms the quality of the global surface shape regarding waviness obtained with the machining process for both 2014 and 2019 mirrors.

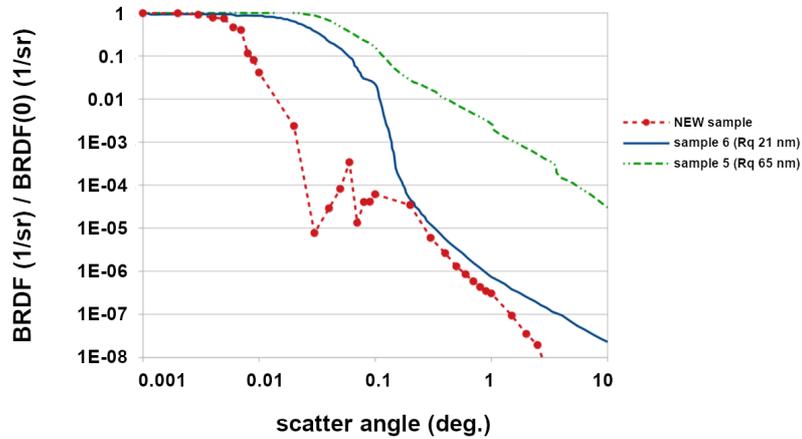

*Figure 7 : Normalized BRDF of mirror witness samples at 350 nm and 5 to 8°*
*incidence angles, showing a significant improvement from old samples (blue and*
*green lines, tested at 8° by CTA-MTF in Czech Republic) to new sample (red line,*
*tested at 5° by Light Tec company).*

The characterization of the recent witness-samples shows a much better micro-roughness below 5 nm rms (measured with an optical profilometer) and a specular reflectivity above 85% in the 300-550 nm wavelength range. Those results were confirmed thanks to Bidirectional Reflectance Distribution Function (BRDF) measurements at different wavelengths, incidence angles and planes of incidence, showing a homogeneous polishing of the surface. A drastic improvement of the local surface quality has been reached relatively to old samples (see Fig. 7).

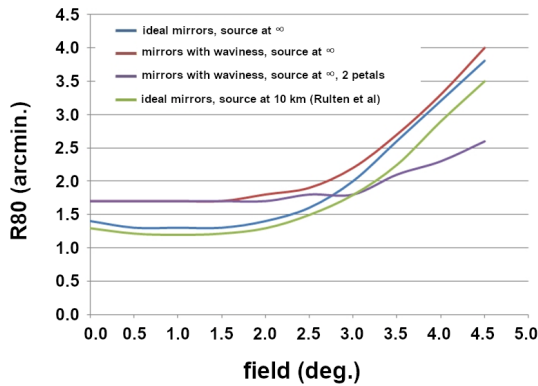

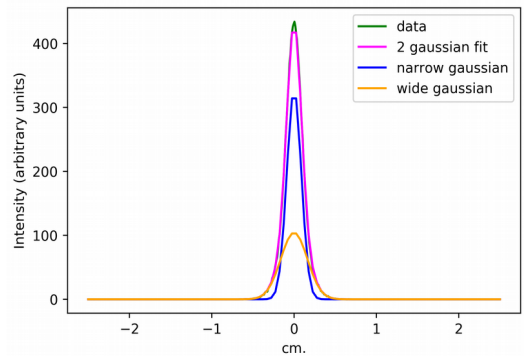

*Figure 8: Specular R80 on and off-axis for prototype mirrors compared to ideal R80 of the GCT theoretical optical system. Deduced from Zemax simulations, M1 and M2 waviness errors included, neglecting micro-roughness and alignment errors. The green curve is taken from [8]*

*Figure 9: Double Gaussian fit of 1D stellar profiles in the case of mirrors with improved roughness of Rq ~ 21 nm, at the wavelength of 325 nm.*

Theoretical and ROBAST extrapolations from first measurements show that improvement in micro-roughness (Rq ~ 21 nm) leads to substantial suppression of the initially observed wide component (cf. Fig. 9), and provides a PSF well within the CTA specifications. Further improvement of Rq to ~ 7 nm even allows to reach a D80 of ~ 2.5 mm on-axis (~0.063°), neglecting alignment errors. Combining FEA (finite element analysis) and ROBAST simulations lead to a final total D80 estimate of ~ 3.3 mm on-axis including misalignment effects.

**5. Conclusions**

   The mechanical, control-command and optical performances of the proposed GCT design are compliant with the CTA requirements. The optical system could even reach significantly better PSF well adapted to finer-pixel cameras with new optimised mirrors, but this still requires validation on sky. This dual mirror S-C design provides multiple advantages over a single-dish construction, in particular, compensation of optical aberrations, large field of view, as well as a compact reliable mechanical structure. The simulations predict that with an improved mirror polishing fairly easily achievable with current technology, the PSF can be significantly better than the CTA specification. Thus, the GCT design shows a great potential for the Cherenkov astronomy in general, and is highly promising for the CTA project.

# References

1.  Vassiliev, V., Fegan, S., Brousseau, P., Astroparticle Phys., 28, 10 (2007)

2.  Vassiliev, V., for the pSCT project, Proc. of 35th ICRC, 301, 35, id838, Busan, Korea (2017)

3.  Dournaux, J.L., et al, The GCT consortium, "Operating performance of the gamma-ray Cherenkov telescope: An end-to-end Schwarzschild-Couder telescope prototype for the Cherenkov Telescope Array," Proceedings of the Vienna conference on instrumentation 2016, Nuclear instruments and methods in Physics research section A, 845, 355-358, (2017)

4.  Sol, H., et al, for the CTA GCT project, Proc. of 35th ICRC, 301, 35, id822, Busan, Korea (2017)

5.  Zorn, J., White, R., Watson, J.J., et al, NimA, 904, 44-63, Characterisation and testing of CHEC-M, a camera prototype for the SST of CTA (2018)

6.  Le Blanc, O., et al., For the CTA GCT Project, "Final characterisation and design of the Gamma-ray Cherenkov Telescope (GCT) for the Cherenkov Telescope Array," Proceedings of SPIE 10700, (2018)

7.  Dournaux, J.L., et al., For the CTA Consortium, "Performance of the Gamma-ray Cherenkov Telescope structure, a dual-mirror telescope prototype for the future Cherenkov Telescope Array," Proceedings of SPIE 9912, (2016)

8.  Rulten, C., Zech, A., Okumura, A., Laporte, P., & Schmoll, J. 2016, Astroparticle Physics, 82, 36

9.  Okumura, A.; Noda, K.; Rulten, C. "ROBAST: Development of a Non-sequential Ray-tracing Simulation Library and its Applications in the Cherenkov Telescope Array". Proceedings of the 34th International Cosmic Ray Conference (ICRC2015). The Hague, The Netherlands.

10.  https://www.zemax.com/products/opticstudio

11.  Dournaux, J.L., Huet, J.M., Amans, J.P., Dumas, D., Laporte, Sol, H., Blake, S., For the CTA Consortium, "Mechanical design of SST-GATE a dual-mirror telescope for the Cherenkov Telescope Array," Proc. SPIE 9151 (2014).

12.  Catalano, Osvaldo; et al. "The ASTRI camera for the Cherenkov Telescope Array". Proceedings of the SPIE, Volume 10702, id. 1070237 16 pp. (2018).